\documentclass[4paper,12pt]{article}
\usepackage[brazilian,british]{babel}
\usepackage[utf8]{inputenc}
\usepackage[T1]{fontenc}
\usepackage{multicol, blindtext, graphicx}
\usepackage{graphicx}
\usepackage{epsfig}
\usepackage[usenames]{color}
\usepackage{amssymb}
\usepackage{amsmath}
\usepackage{array}
\usepackage{tabularx}
\usepackage{cite}
\usepackage{times}
\usepackage{authblk}
\usepackage{setspace}
\usepackage[left=2cm,top=3cm,right=1.5cm,left=1.5cm,foot=2.5cm]{geometry}
\usepackage[runin]{abstract}
\usepackage[compatibility=false,labelfont=it,textfont={bf,it}]{caption}
\abslabeldelim{:\,}

\title{Analyzing course programmes using complex networks}
\author{Suzane F. Pinto}
\affil{\small\textit{Programa de graduação em Engenharia de Computação,
Universidade Federal de Ouro Preto, 35931-008, Brasil}}
\author{Ronan S. Ferreira}
\affil{\small\textit{Departamento de Ciências Exatas e Aplicadas,
Universidade Federal de Ouro Preto, 35931-008,  Brasil}}
\date{}
\newenvironment{multiabstract}[1]
  {\renewcommand{\abstractname}{#1}\begin{abstract}}
  {\end{abstract}}
\begin{document}
\maketitle
    \maketitle
\selectlanguage{british}
    \begin{multiabstract}{Abstract}
We analyze the curriculum of the early common-years of engineering in our institute using tools of statistical physics of complex networks. Naturally, a course programme is structured in a networked form (temporal dependency and prerequisites). In this approach, each topic within each programme is associated with a node, which in turn is joined by links representing the dependence of a topic for the understanding of another in a different discipline. As a course programme is a time-dependent structure, we propose a simple model to assign links between nodes, taking into account only two ingredients of the teaching-learning process: recursiveness and accumulation of knowledge. Since we already know the programmes, our objective is to verify if the proposed model is able to capture their particularities and to identify implications of different sequencing on the student learning in the early years of engineering degrees. Our model can be used as a systematic tool assisting the construction of a more interdisciplinary curriculum, articulating between disciplines of the undergraduate early-years in exact sciences.\\
\textbf{Keywords}: course programme, learning, complex systems.
\end{multiabstract} 
%    \begin{multiabstract}{Resumo}
%Analisamos as grades curriculares do ciclo básico das engenharias em nosso instituto usando ferramentas da física estatística de redes complexas. Naturalmente, uma grade curricular estrutura-se em forma de rede (ordenamento temporal e pré-requisitos). Nessa abordagem, cada tópico dentro de uma ementa é associado a um nó que, por sua vez, são conectados por links que representam a dependência de um certo tema para a compreensão de um outro, em uma disciplina diferente. Como uma grade curricular é uma estrutura tempo-dependente, evoluindo semestre após semestre, propomos um modelo simples para assinalar ligações entre pares de nós levando em conta apenas dois ingredientes do processo de ensino-aprendizagem: a recursividade e o acúmulo de conhecimentos. Como já conhecemos nossas grades curriculares, nosso primeiro objetivo é verificar se o modelo proposto é capaz de capturar as particularidades de cada uma delas e identificar as implicações que diferentes sequenciamentos do setor de física possam ter no aprendizado dos estudantes. Nosso modelo pode ser usado como uma ferramenta sistemática auxiliando na construção de uma grade curricular interdisciplinar, articulando entre os saberes das disciplinas iniciais da graduação em ciências exatas.\\
%\textbf{Palavras-chave}: grade curricular, ensino, sistemas complexos.
%    \end{multiabstract}
\singlespacing

\begin{multicols}{2}
\section{Introduction}
\label{sec:intro}
The theory of complex networks has its inaugural landmark in the late 1990s ~\cite{WS98,barabasi1999emergence}. Since then, several areas of knowledge have passed through~\cite{barabasi:takeover,newman2010networks,barabasi2017elegant}. From investigations into social dynamics~\cite{PhysRevE.97.012305},through spread and control of epidemics~\cite{masuda2017temporal,pastor2018eigenvector,PhysRevLett.120.068302} and even linguistics~\cite{Martin2016117,torre2017emergence}. In this way, the theory of complete networks has become a paradigm for interdisciplinary research that, in turn, is a challenge for teaching, production and technical-scientific dissemination. However, our curriculum is divided into specific content, due to a didactic motivation, taking into account the teaching and learning process. Another point is the prerequisites: requirements on skills acquired for continuity and the accumulation of knowledge. Thus, the prerequisites are, at least in the context of interdisciplinarity, a connection mechanism between the various disciplines offered in a curriculum.

On the other hand, the knowledge of a larger fraction of the contents in a curricular grid by the teachers is also a fundamental mechanism for the consolidation of the connections between the disciplines. This is because, on the part of the students, it is not always easy to perceive the relevance of the topics of one discipline in another. Perhaps for this reason, questions such as:  ``\emph{When will I use this?}'' or ``\emph{Why am I learning this?}'' accompany the student's daily life.

Several studies aim to improve the question of integrating the contents of the basic cycle ~\cite{belanccon2017ensino,matta2017}. Particularly, in engineering undergraduate courses, the failure rates in the disciplines of this stage have been transformed into a kind of “culture of failure”. This thought leads \emph{freshmen} to consider the repetition of subjects such as \emph{Differential and Integral Calculus} and \emph{Physics} a natural fact: the rule, the exception being that regular student in the course in question \cite{barroso2004evasao,belanccon2017ensino,matta2017}. Consequently, retentions lead to a high number of dropouts in some courses, implying a large number of idle vacancies that are difficult to be reoccupied when in more advanced periods of the courses \cite{campello2008metodologia}.

In our institution, the Federal University of Ouro Preto - UFOP, the Pro-active program was created, which aims to improve the offer conditions of undergraduate courses and disciplines and the learning process. This is encouraged with the participation of students, who are selected every year by the program. Our project proposed the exposure of selected students to the theory of complex networks. From there, the group was encouraged to map the disciplines of the basic cycle, building a network of themes connecting the various disciplines.

A network is a graph ~\cite{newman2010networks} in which we assign a physical meaning to a set of nodes (vertices) connected by edges, obeying some statistical distribution. In our approach, each topic within each menu of the disciplines of the basic cycle is associated with a node,which in turn are connected in pairs by edges (links). These links represent, therefore, the dependence of a certain theme in one discipline for the understanding of another in a different discipline. In this way, we can build a network of themes from the connections between the topics of the menus. We use the menus of the basic cycle curriculum offered at the Institute of Exact and Applied Sciences - ICEA - UFOP.

Our objectives then are (\emph{i}) the proposal of a model to point out the dependencies between the different themes of a curricular menu;(\emph{ii}) check if the proposed model is capable of capturing the particularities of each studied grid. Verification is possible because we know the grids beforehand; Finally, (\emph{iii}) identify from the model the influence of the different sequences of physics themes both for the curriculum and for the teaching-learning process.

\section{Methodology}
We can use adjacency matrices to represent graphs. For a graph of type  $G(N,E)$, where $N$ is the number of nodes and $E$ the number of edges, we can write a matrix of type  $A:N\times N$ naming each node with an integer $i=1,2,3,...,N$. The adjacency matrix carries information about the existence of a connection between any two nodes in a graph. For a directed type graph, the  $\{a_{ij}\}$ elements of matrix  $A$ are defined as follows:
\[
a_{ij}=
 \left\{
	\begin{array}{lll}
		1,\textrm{if an $i$ connection}\\
	  \textrm{for $j$ already exists};\\\\
		0,\textrm{ if an $i$ connection }\\	 
	  \textrm{for $j$ no longer exists}.
	\end{array}
\right.
\]\\
A targeted graph can, for example, represent a social network such as \emph{Facebook} or \emph{Instagram}: we don't always follow people who know us, as is the case with celebrity followers. This can be understood as a connection that preserves meaning. So we can represent that connection with a arrow indicating a direction in the link between the celebrity and the unknown fan. On the other hand, in the case of an undirected graph, we can write the set of elements of a matrix $B:N\times N$ from its  $\{b_{ij}\}$ entries, where
\[
b_{ij}=
 \left\{
	\begin{array}{lll}
		1,\textrm{ if a connection between $i$}\\
	 \textrm{and $j$ already exists};\\\\
		0,\textrm{if a connection between $i$}\\
		\textrm{and $j$ does not exist}.
	\end{array}
\right.
\]\\
The diagonal of an adjacency matrix guards the entry of elements $\{a_{ii}\}$ (or $\{b_{ii}\}$),
informing if a node is connected to itself. If that happens, he has a self-connection. In general, $a_{ij}\neq a_{ji}$, but $b_{ij}=b_{ji}$
always, so $B$ is a symmetric matrix.

The most fundamental of the quantities studied in networks is the degree of a node, given by the number of connections it has. The distribution of degrees $p(k)$ in a graph is equivalent to the fraction of nodes that have degree $k$:
\begin{equation}
 p(k)=\frac{N_{k}}{N},
\end{equation}
since $\sum_k N_{k}=N$. The $k_{i}$ degree of a node $i$ can be obtained from the adjacency matrix of a graph $G(N,E)$. Assuming undirected $G$, then
\begin{equation}
  k_{i}=\sum_{j=1}^{N} b_{ij}.
\end{equation}
\\
If it is directed, we will have to count the number of edges that enter (arrow pointing to node $i$),
$k_{i}^{in}$, and that come, $k_{i}^{out}$, in an node
$i$:
\begin{equation}
  k_{i}^{in}=\sum_{j=1}^{N}a_{ji}
\end{equation}
e
\begin{equation}
  k_{i}^{out}=\sum_{j=1}^{N}a_{ij}.
\end{equation}

The average value of degrees $\langle k\rangle$ that a graph can cause from the distribution of connectivity $p(k)$. For example, for an undirected graph, the average value  $\langle k\rangle$ can be written as: 
\begin{equation}
\langle k\rangle = \frac{1}{N}\sum_{i}k_{i}=\sum_k kP(k).
\end{equation}

\section{Model}
\label{sec:model}
By mapping the course menus in an adjacency matrix, we can associate a node with each topic and establish a link representing the correspondence of a given topic in a certain discipline to understand the content in another. In this way, we obtained a targeted type network, in which the links have their ends with arrows. In addition, our network also has a temporal dependency, because, of course, disciplines from later semesters may depend on content presented in past semesters. 
Therefore, our network of themes grows with each period and its links have a direction.

In practice, in order to deal with the time-dependent character between nodes and build our network of themes, we propose the following model:
\begin{itemize}
\item[(\emph{i})] Topics in the same period are allowed to be connected in a distributive manner;
\item[(\emph{ii})] Topics in a period $\wp$ can also connect with the period $\wp-1$ (com $\wp>2$), thus giving the ingredient of recursion, portraying the idea of the accumulation of knowledge.
\end{itemize}

\begin{figure*}[ht]
\centering
\includegraphics[width=4.5cm]{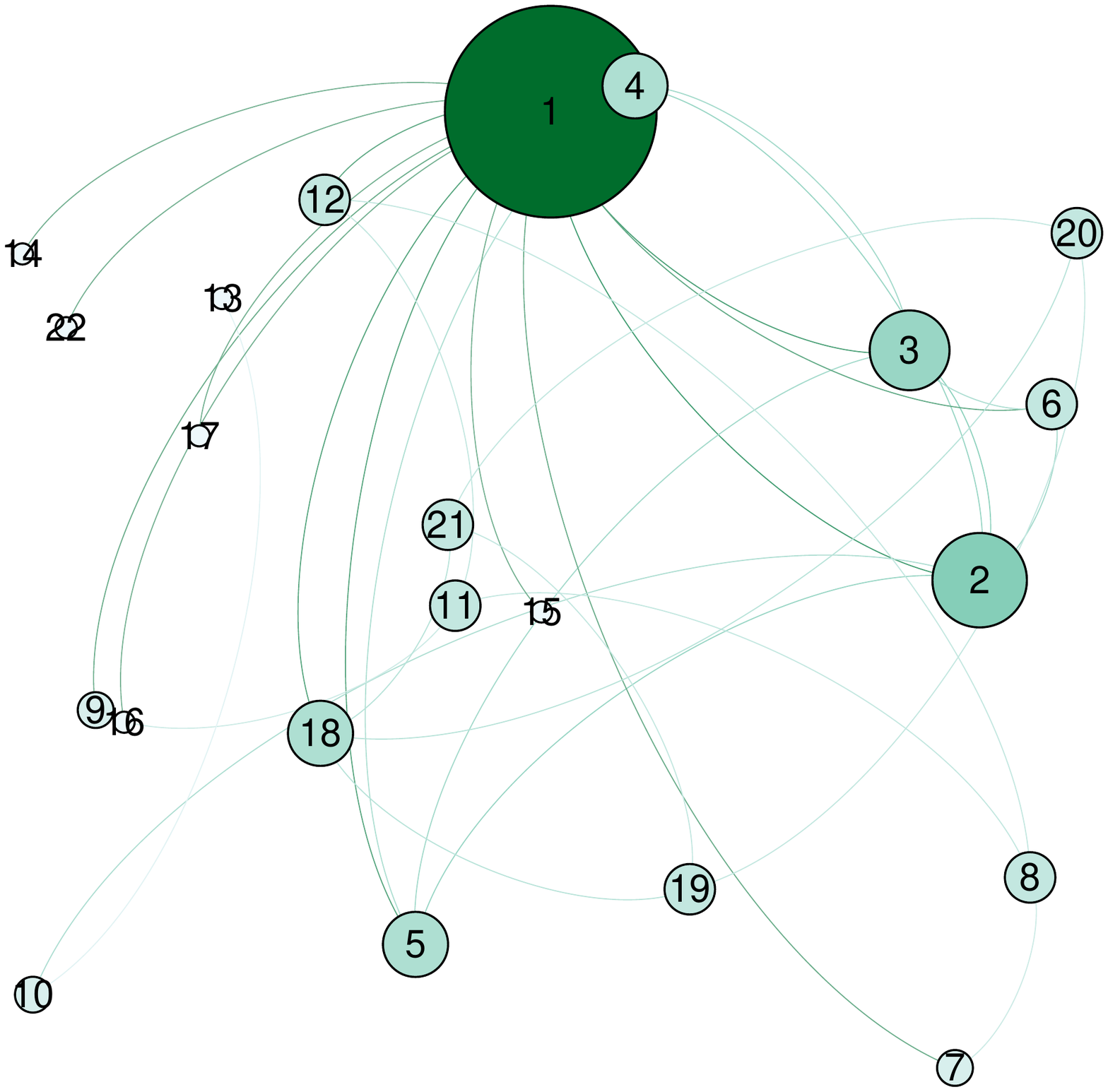}\includegraphics[width=4.5cm]{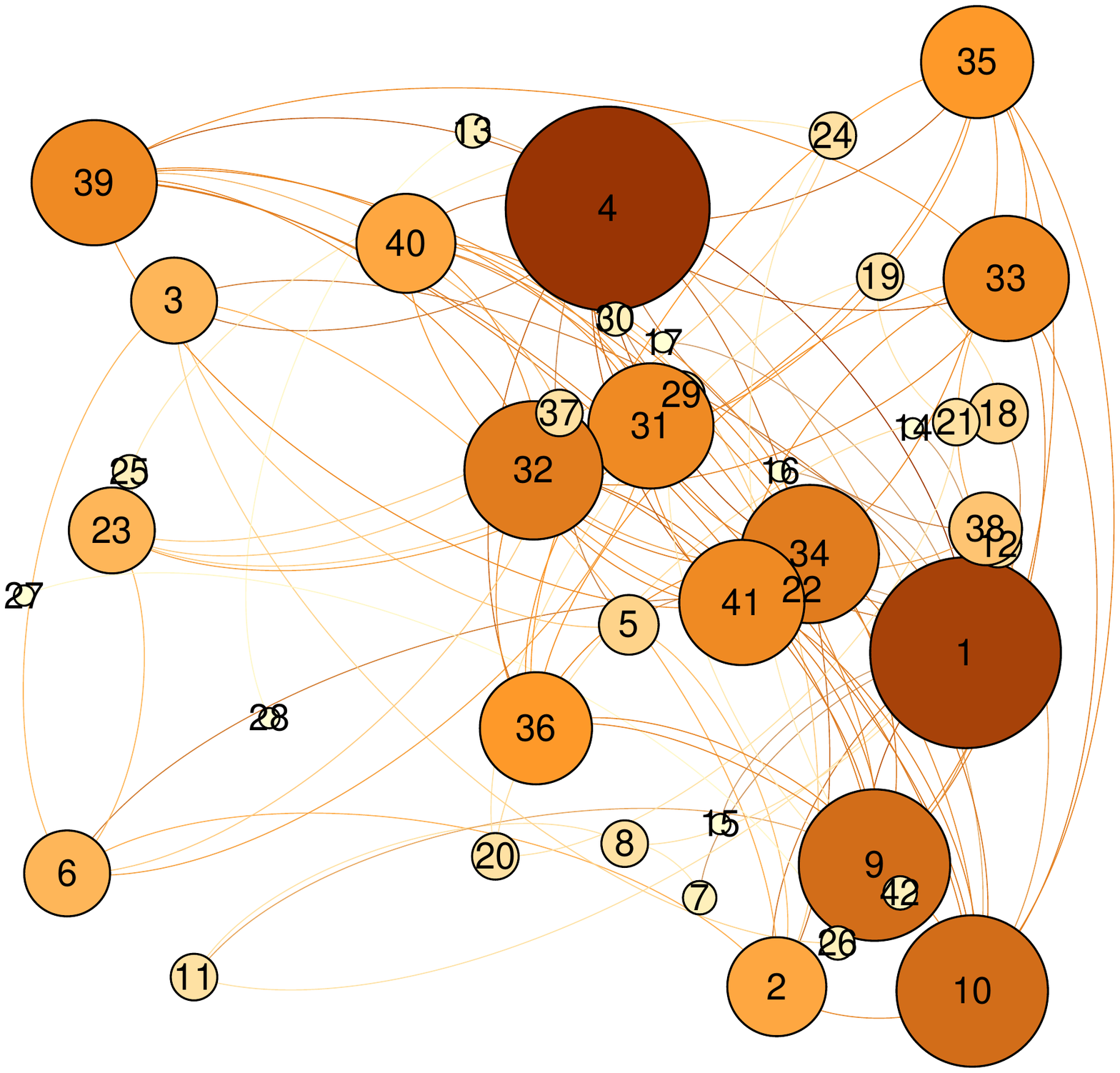}
\includegraphics[width=4.5cm]{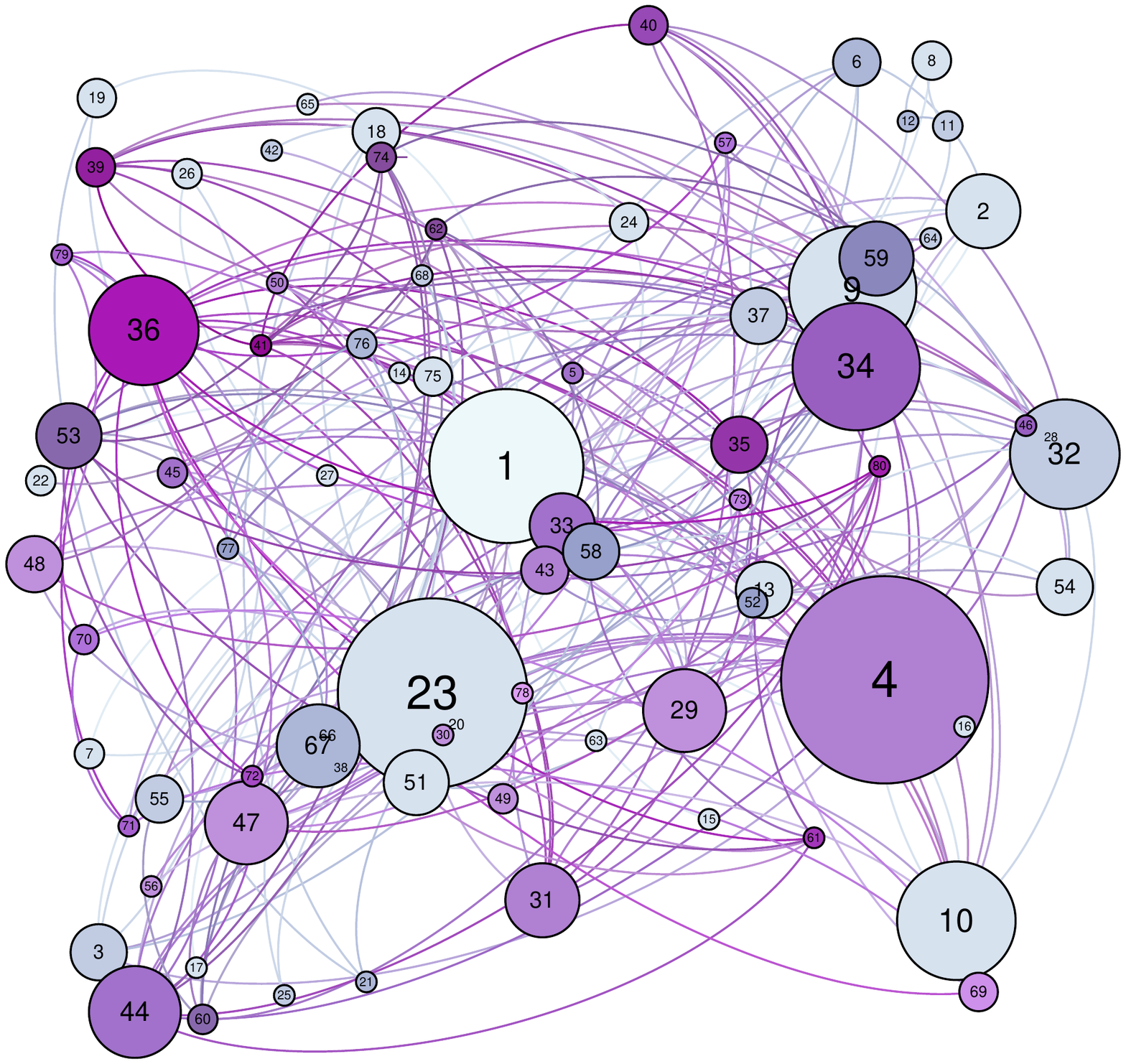}\includegraphics[width=4.5cm]{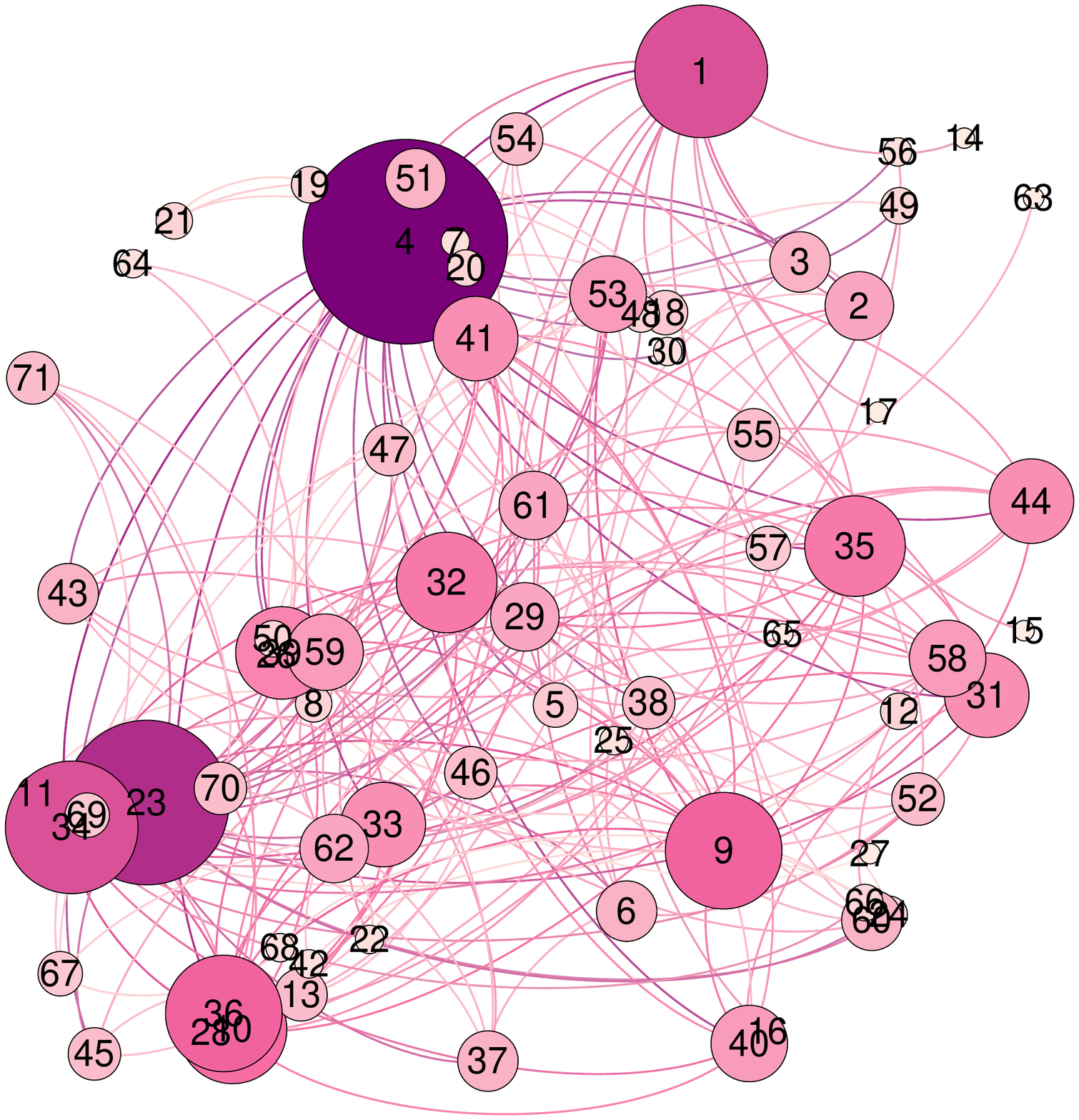}
\caption{Illustration of the temporal character for increasing the density of correlations and corresponding number of nodes to the courses offered in the 1st Period (top, left) and clockwise for the 2nd, 3rd and 4th Periods of the Electrical Engineering course / UFOP.}
\label{fig:rede}
\end{figure*}

Figure \ref{fig:rede} illustrates this mapping of menu topics in an undirected network. Each topic is named by a number\footnote{This correspondence is not shown for brevity.}.

\section{Results and discussions}
The distributions that we found show too many fluctuations, which makes the analysis for statistical purposes difficult. These fluctuations are mainly at the tail of distributions where $x>>1$ or, equivalently, for low frequencies of occurrence of $x$. In order to carry out a statistical treatment in $p(k)<<1$ values, one strategy is to use \textit{complementary cumulative  distribution functions} - CCDF. In  general way, a CCDF function is such that, at the continuous limit, $f(x)=1-\int_{-\infty}^xg(x')dx'$, where $x$ is a variable  random  corresponding to the variable, also random, $x'$, truncated in the interval  $[-\infty,x]$.For our the discreet case suffices, in which we will denote a function of this type by  $p_>(k)$:
\begin{equation}
p_>(k)=\sum_{q\geq k}p(q).
\end{equation}

\begin{figure*}[ht]
\centering
\includegraphics[width=8.25cm]{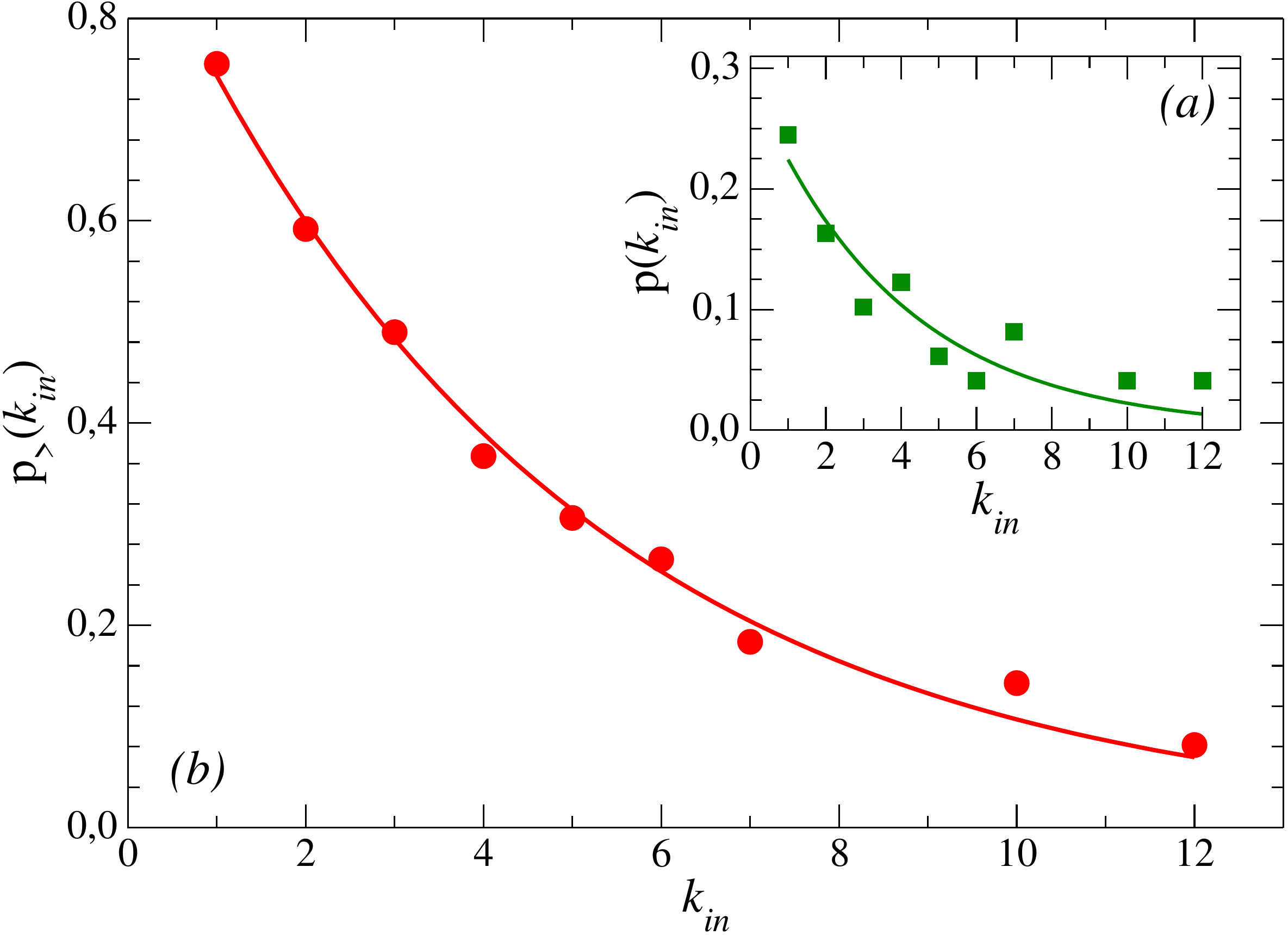}
\includegraphics[width=8.25cm]{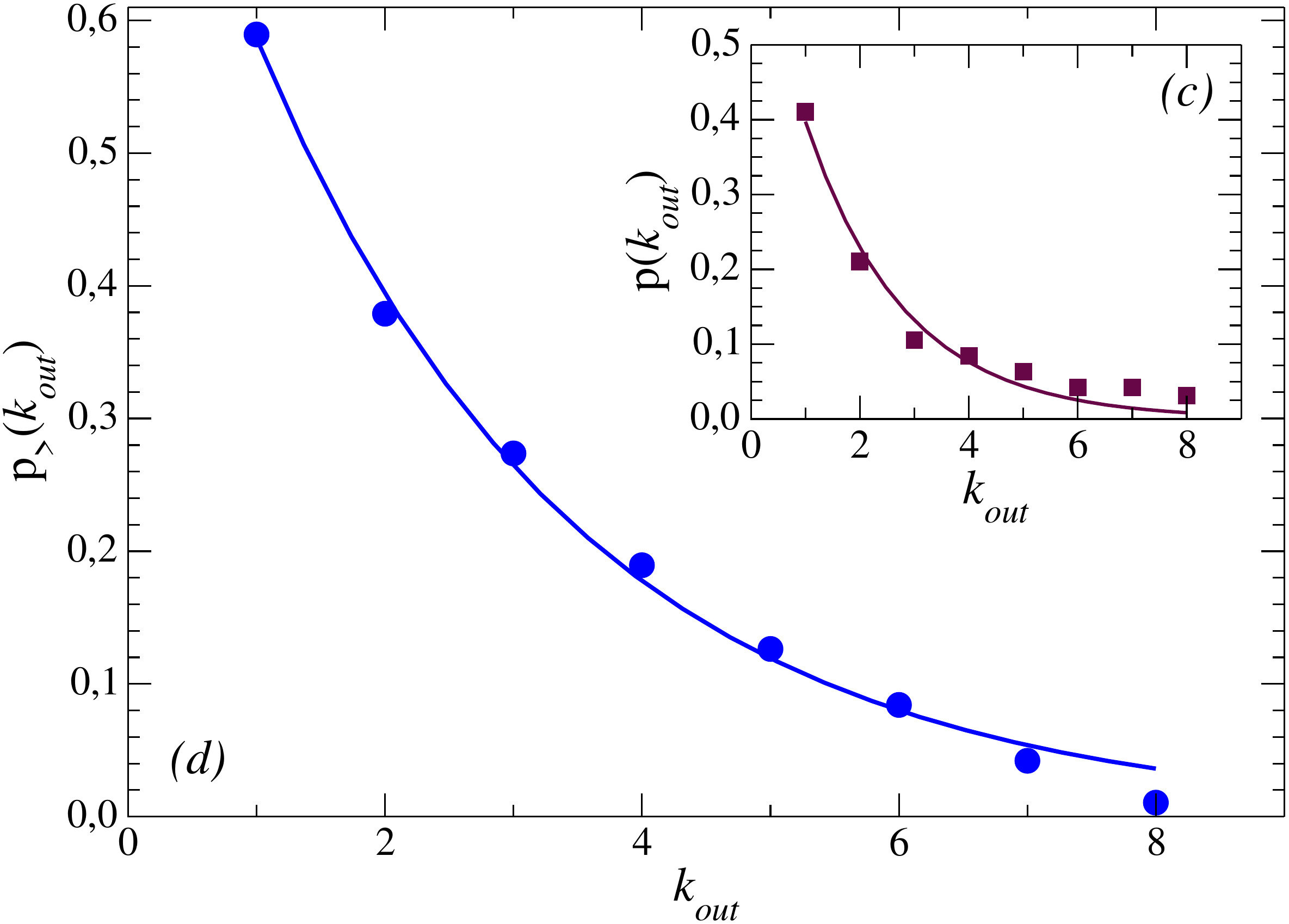}
\caption{From the inserted graphs \textit{(a)} and \textit{(c)} to the respective main graphs \textit{(b)} and \textit{(d)}, the data suffered a treatment due to fluctuations for $p(k)<<1$. Left: degree distributions for incoming links. Right: degree distributions for outbound links.}
\label{fig:1}
\end{figure*}

At the top of figure~\ref{fig:1} the input distributions $(a)$ $p(k_{in})$ and (b) $p_>(k_{in})$, are shown, while at the bottom the output distributions $(c)$ $p(k_{out})$ and $p_>(k_{out})$ are shown for the Electrical Engineering (EE) themes network. In \textit{(a)}, the data refer to the distribution without the CCDF statistical treatment, which is shown in  \textit{(b)}. The best fit found, following the proportionality $p(k)\propto \exp{(-\alpha k)}$ is shown, both in \textit{(a)} as in \textit{(b)}, by the continuous plot curves. The same sequence of information is presented in \textit{(c)} and \textit{(d)}for outbound distributions. It is interesting, at this point, to make an observation about the process of building the network of themes. Our first investigation was just to connect those themes related, without any concern with the temporal dependence nor thinking about the process of accumulation of knowledge, on which the learning process is essential based. In this case, what we found was an exclusively random distribution of points randomly ordered on the $xy$ plane (data not shown). Only with our model proposal presented in section~\ref{sec:model} which is obtained as distributions prohibited in figure~\ref{fig:1}. 

\begin{figure*}[!htbp]
\centering
\includegraphics[width=16.4cm]{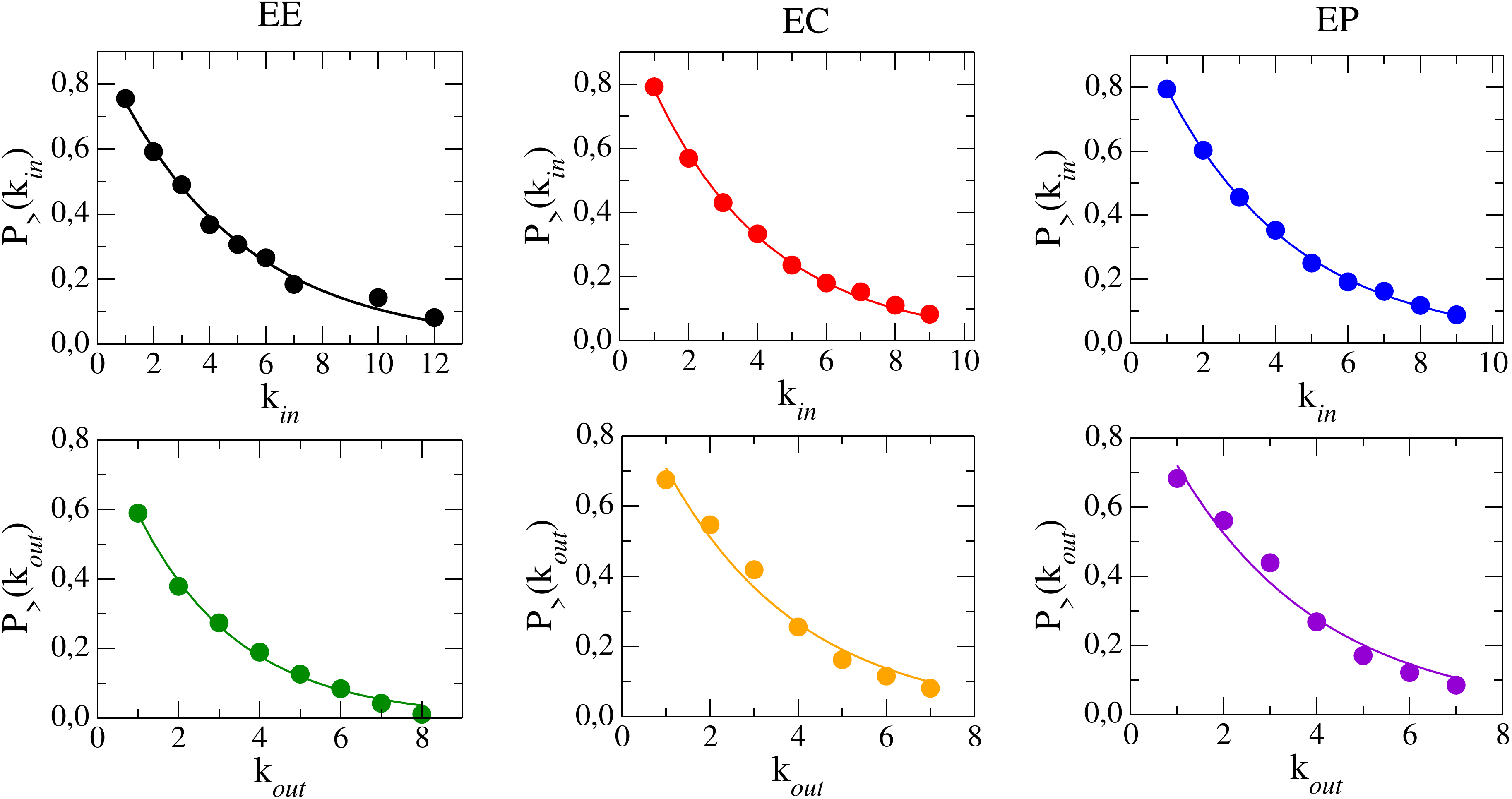}
\caption{$p_>(k_{in})$ distributions (top) and 
$p_>(k_{out})$ > (kout) distributions (base) for connections between the topics of menus of the three studied engineers. from left to right: Electrical Engineering (EE), Engineering Computing (EC) and Production Engineering (EP). The values for exponents of the adjustment for each of the curves is shown in table ~\ref{tab:dat}.}
\label{fig:2}
\end{figure*}

Figure~\ref{fig:2} shows the input and output distributions for the three UFOP engineers analyzed: Electrical Engineering (EE), Computer Engineering (EC) and Production Engineering (EP) The trace curves are exponential adjustments and the values of the respective exponents are shown in table~\ref{tab:dat}.
\begin{table*}[!htbp]
 \centering
\begin{tabularx}{1.0\textwidth} { 
   >{\raggedright\arraybackslash}X 
  | >{\centering\arraybackslash}X 
  | >{\centering\arraybackslash}X 
  | >{\centering\arraybackslash}X 
  }
 \hline\hline
& EE & EP & EC\\ \hline
Nodes & 96 & 87 & 96\\
links & 259 & 293 & 300\\
$\langle k_{in}\rangle$ & 5,22(1) & 4,28(1) & 4,14(1)\\
$\langle k_{out}\rangle$ & 2,69(1) & 3,55(1) & 3,47(1)\\
$\alpha_{in}$ & 0,22(1) & 0,28(1) & 0,29(1)\\
$\alpha_{out}$ & 0,39(1) & 0,32(1) & 0,33(1)\\\hline
 \end{tabularx}
 \caption{Quantitative characterization of the engineering themes network. Correspondence: EE (Engineering Electrical), EP (Production Engineering) and EC (Computer Engineering).}
 \label{tab:dat}
\end{table*}

\begin{figure*}[!htbp]
\centering
\includegraphics[width=9.0cm]{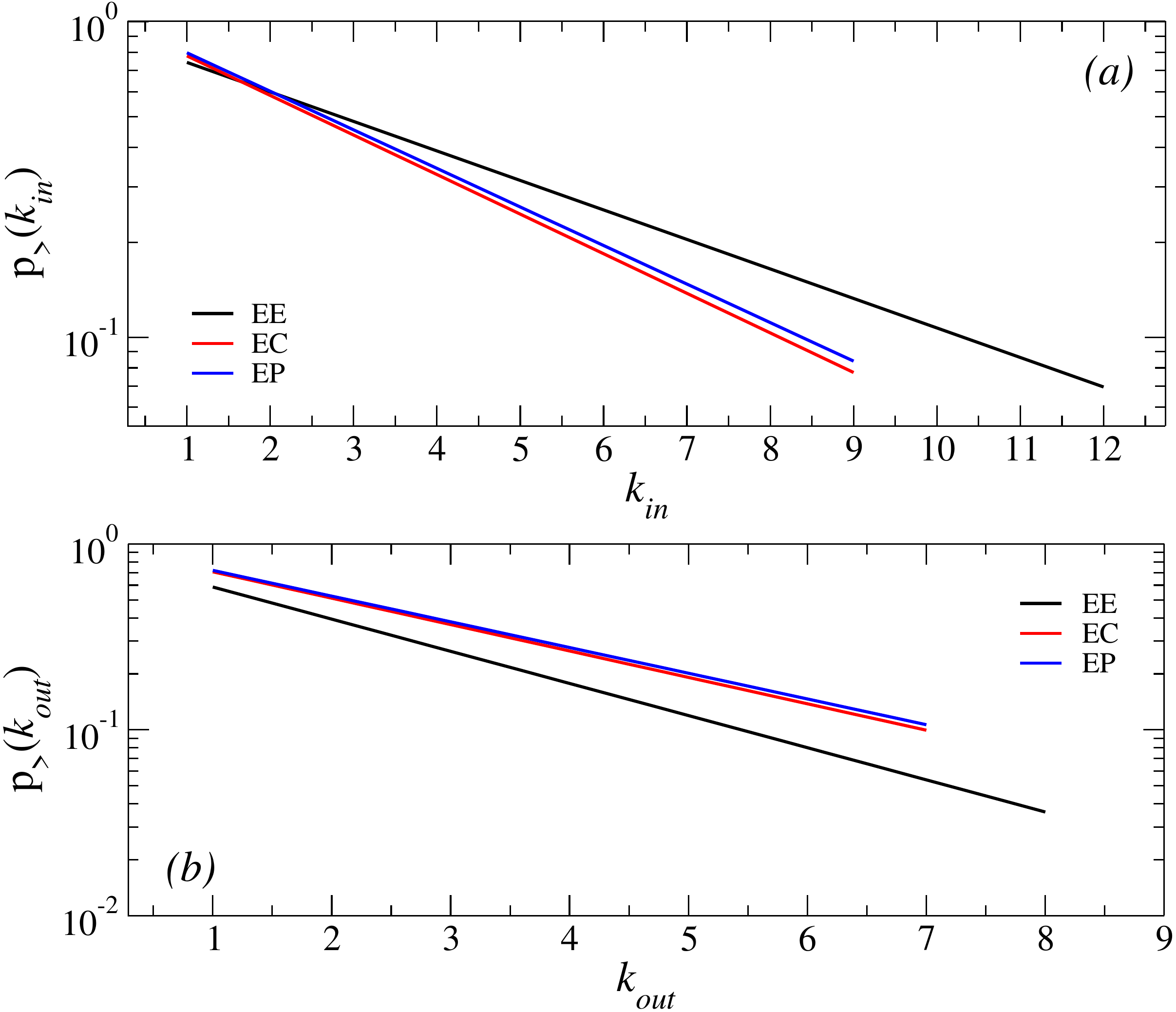}
\caption{Comparison between the continuous functions obtained in figure ~\ref{fig:2}. How is the network of themes for the basic cycle of three engineering companies was expected to overlap curves. However, this does not occur due to differences in the Pedagogical Curricular Program of the courses. Our model was able to capture these differences using a quantitative approach.}
\label{fig:3} 
\end{figure*}

For comparison, the continuous functions obtained from the exponential adjustments shown in figure~\ref{fig:2} are placed on the same scale in figure ~\ref{fig:3}. Note that, although it is data from the basic cycle of Engineering and, therefore, we hope to obtain a superposition
of the inclinations of all the studied Engineering, this it only happens with data related to Computer and Production Engineering, respectively, EC and EP. The superposition is not perfect due to slight differences in the menus of the two courses. The most prominent of these differences lies in the fact that the EC has in its PPC the discipline of \emph{Modern Physics (or Physics vol. IV)}, which does not happens for EP.

This suggests that the inclusion of Physics IV in the EP curriculum would be a proposal to the PPC course, in order to offer students a current view of the subjects covered, implying a better understanding of processes in modern engineering, with emphasis on the processes of physics and materials engineering.

Another interpretation of figure 3 is that the inclusion of Modern Physics in the EP curriculum would not impose an overload on students from the point of view of accumulation of knowledge, number of prerequisites that would be required for such discipline. Otherwise, the deviation between the slopes of the EC and EP curves would be evident, both for the distribution of input connections and for connections of output, since the number of incoming connections of a node is related to the number of prerequisites for the theme that this node represents, while its number of outgoing connections is associated with the relevance of this theme for the course progress.

\begin{table*}[ht]
\centering
\begin{tabularx}{1.0\textwidth} { 
   >{\raggedright\arraybackslash}X 
  | >{\centering\arraybackslash}X 
  | >{\centering\arraybackslash}X 
  | >{\centering\arraybackslash}X 
  }
\hline\hline
Period & EE & EC & EP\\ \hline
$\wp=2$ & FIS I & FIS I & FIS I\\
$\wp=3$ & FIS II \& FIS III & FIS II & FIS II\\
$\wp=4$ & FIS IV & FIS III & FIS III\\
$\wp=5$ & - & FIS IV & -\\\hline
\end{tabularx}
\caption{Correspondence: FIS I (Mechanics), FIS II (Electromagnetism),
FIS III (Thermodynamics) e FIS IV (Modern Physics).}
\label{tab:fis}
\end{table*}

The most substantial deviation occurs between Electrical Engineering, EE, and the others. In this case, the superposition doesn't happen because there are also particularities in its pedagogical project. Again, it is the physics sector that presents the main differences. For example, in EE's PPC the discipline of \emph{Electromagnetism} (or Physics vol. III) and \emph{Thermodynamics} (or Physics vol. II)) are in the same semester, while separated by a period in the other EC and EP.

Another factor is that the discipline of \emph{Modern Physics} at EE is in the semester immediately after ($\wp$ and $\wp+1$) to the period in which the discipline of \emph{Electromagnetism} is programmed. This is because in its \emph{Modern Physics} menu the first chapters are dedicated to the study of \emph{alternating currents}, which in turn, have direct relationship with the study of \emph{circuits} in the discipline \emph{Electromagnetism}. In our approach, the nodes that represent the theme of \emph{alternating currents}, in EE, have $k_{out}$ values higher than the same nodes in the EC network, because in this engineering \emph{Electromagnetism} and \emph{Modern Physics} are separated from a period. 

The reverse reasoning applies to the $k_{in}$ values for the nodes that represent the study of the \emph{circuits}. The table~\ref{tab:fis} outlines the positioning of these disciplines in the curriculum grids of the three engineering companies. These factors lead to a higher average connectivity for EE showing a richer network in prerequisites - curve with greater slope in figure  3  (a),  although an unbalanced grid if we take into account the relationship between prerequisites and relevance of a theme following a course. This is related to the ratio between the coefficients of the input and output connection distributions, where this ratio $r$ tends to 1 if there is a linear correlation. Namely,  $r=0.6$ for EE, $r=0.8$ for EP  and $r=0.9$ for EC, the latter showing the course here analyzed with the most balanced grid from the point of view of the sequence of presentation of the themes of teaching of physical.

\section{Final comments}
Within the scope of the Pro-Ativa / UFOP project, which aims to think about learning and teaching in undergraduate courses, we a simple model assuming in the learning process the ingredients of recursion and accumulation knowledge. To test our model, we studied the curricula of the basic cycle of the three engineering offered at ICEA / UFOP. With mapping of correspondences between grid topics curriculum in a network of themes, we were able to obtain mathematical distributions describing each of these
grids. Although we only deal with the basic cycle of these courses, the PPC of one of these engineering differences compared to the others. Our model was able to capture these differences quantitatively while treat the connectivity distributions for each of the curricular grids. Our analysis can be extended both to the professional cycle and to assist in the study of prerequisites in a PPC.

\section{Acknowledgments}
The authors would like to thank the Pro-Active Program / UFOP. RSF would like to thank the Auxílio Pesquisador program / UFOP, to professors from DECEA / UFOP, UESPI-Piripiri-PI / Teresina-PI, DFIS / UFPI and National Council of Scientific and Technological Development - CNPq, in the scope of process 424950 / 2018-9.

\end{multicols}
\end{document}